# Fracture-Driven Single Bubble Grows and Migration Model in Aquatic Muds
(method article)

Regina Katsman*

The Dr. Moses Strauss Department of Marine Geosciences, Leon H. Charney School of Marine Sciences, The University of Haifa, Haifa, Israel

**\*Corresponding author's email address**
rkatsman@univ.haifa.ac.il

**ABSTRACT**

Methane ($CH_4$) is the most prevalent hydrocarbon and a significant greenhouse gas found in the atmosphere. Buoyancy-driven $CH_4$ bubble growth and migration within muddy aquatic sediments are closely associated with sediment fracturing. This paper presents a model of buoyancy-driven $CH_4$ single bubble growth in fine-grained cohesive (muddy) aquatic sediment.

- Solid mechanics model component simulates bubble elastic expansion caused by solute supply from the surrounding mud, followed by differential fracturing of the mud by the evolving bubble front, a process governed by the principles of Linear Elastic Fracture Mechanics (LEFM). This differential fracturing controls the evolving shape and size of the bubble.

- The model integrates the LEFM with the dynamics of solute exchange between the bubble and the surrounding mud, alongside the conservation of $CH_4$ gas within the bubble.

- An advanced meshing strategy allows balancing between the geometry resolution and the amount of mesh elements, thereby optimizing for both solution accuracy and computational efficiency.

This model is intended to be a foundational tool for proper upscaling of single bubble characteristics to effective gassy medium theories. This will enhance the accuracy of the acoustic applications and could contribute to evaluation of overall $CH_4$ emission from the aquatic muds.

**Graphical abstract**

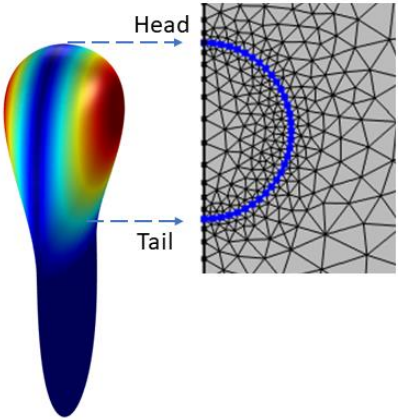

1. Introduction

Gas bubbles found in shallow muddy aquatic sediments pose a significant threat to environment safety, climate sustainability and mechanical stability of the sediment. These bubbles are inclusions of immiscible gas, primarily methane ($CH_4$), which are generated by microbial decomposition of organic matter present in sediment [1]. $CH_4$ gas bubble nucleates under supersaturated conditions in muds. Initially, the bubble grows below the pore scale consuming ambient dissolved $CH_4$ until it entirely fills a single pore space [1,2]. As the bubble continues to expand, it expels the pore water and deforms the muddy sediment skeleton, pushing aside the individual grains to grow beyond the confines of a single pore scale [2,3]. The largest bubbles in muds are typically isolated, non-spherical, disc-shaped cavities with low aspect ratio, capable of reaching over a centimeter in their maximum dimension, surrounded by saturated sediment [1,4].

The growth and subsequent rise of $CH_4$ bubbles within fine-grained aquatic muds are governed by a system of two-way coupled mechanical and biogeochemical processes [1]. Experimental studies conducted using various material, including natural muds [3,5,6], sludge [2,7], and even artificial substitutes like gelatin [3], and magnesium hydroxide ($Mg(OH)_2$) soft sediment [8], consistently demonstrate that that mechanical behavior of these materials in response to bubble growth beyond the pore scale can be accurately described by the principles of Linear Elastic Fracture Mechanics (LEFM) theory [9,10]. Consolidated soft clay-bearing (cohesive) fine-grained water-saturated sediment (mud) behaves mechanically as a single-phase medium with a negligible pore pressure effect [1,11,12]. The principles of LEFM theory are particularly applicable to muds that have been consolidated to high shear strength (*$c_u$>1 kPa* [8]) but exhibit low fracture toughness (*$K_{Ic} \cong 100 Pa \cdot m^{1/2}$* [8,10,13]). Under these conditions, the zone of plastic deformation occurring at the tip of a crack/bubble is consistently smaller compared to the crack size [9,14], making LEFM a valid and effective theoretical framework.

CH$_4$ bubble growth beyond the pore scale in consolidated muds, involves a two-stage process: initial elastic cavity expansion, followed by localized fracturing of the sediment [7,8,15]. This fracturing occurs because the high capillary-entry pressure in fine-grained muds prevents the non-wetting gas phase from entering the new pore throats without breaking the existing inter-particle bonds [16]. This fracturing mechanism is energetically more favorable compared to capillary invasion that has been unequivocally supported by both experimental [10,17] and theoretical [3,16,18-21] evidence. Consequently, the LEFM theory effectively characterizes large sediment-displacing CH$_4$ bubbles in muds as Mode I (tensile) gas-filled cracks [5,20], and is proven to accurately predict their shapes, sizes, and orientations within the muddy sediment [4,6].

## 2. Method details

The modelling of bubble growth by fracture within soft aquatic muds is built upon foundational approaches for modelling bubble growth within an elementary cell, initially proposed for fluids [22-24], and subsequently adapted for muds [18,19]. This sophisticated mechanical-reaction-transport model effectively integrates two processes: CH$_4$(g) gas bubble accumulation which is fed by the diffusion of dissolved CH$_4$(aq) from the surrounding pore waters, and a Linear Elastic Fracture Mechanics (LEFM) component that accounts for the sediment mechanical response. This approach was further developed in [12,20,21,25-29].

The geometric setup for the modelling is illustrated in Figure 1. Sediment cell is presented as a block cut by a symmetry plane (only half of the block is modeled). The simulation of bubble growth commences with a 3D penny-shaped macroscopic seed. This seed is embedded at a symmetry plane of a sediment box, as shown in Figure. 1. Bubble growth is modelled till the development of a "mature" bubble configuration resembling an inverted tear drop with a closed tail in its main cross-section (for example, refer to Figure 3 below, and also [12,20,26]). This specific configuration signifies the point at which the bubble begins its ascent towards the seafloor.

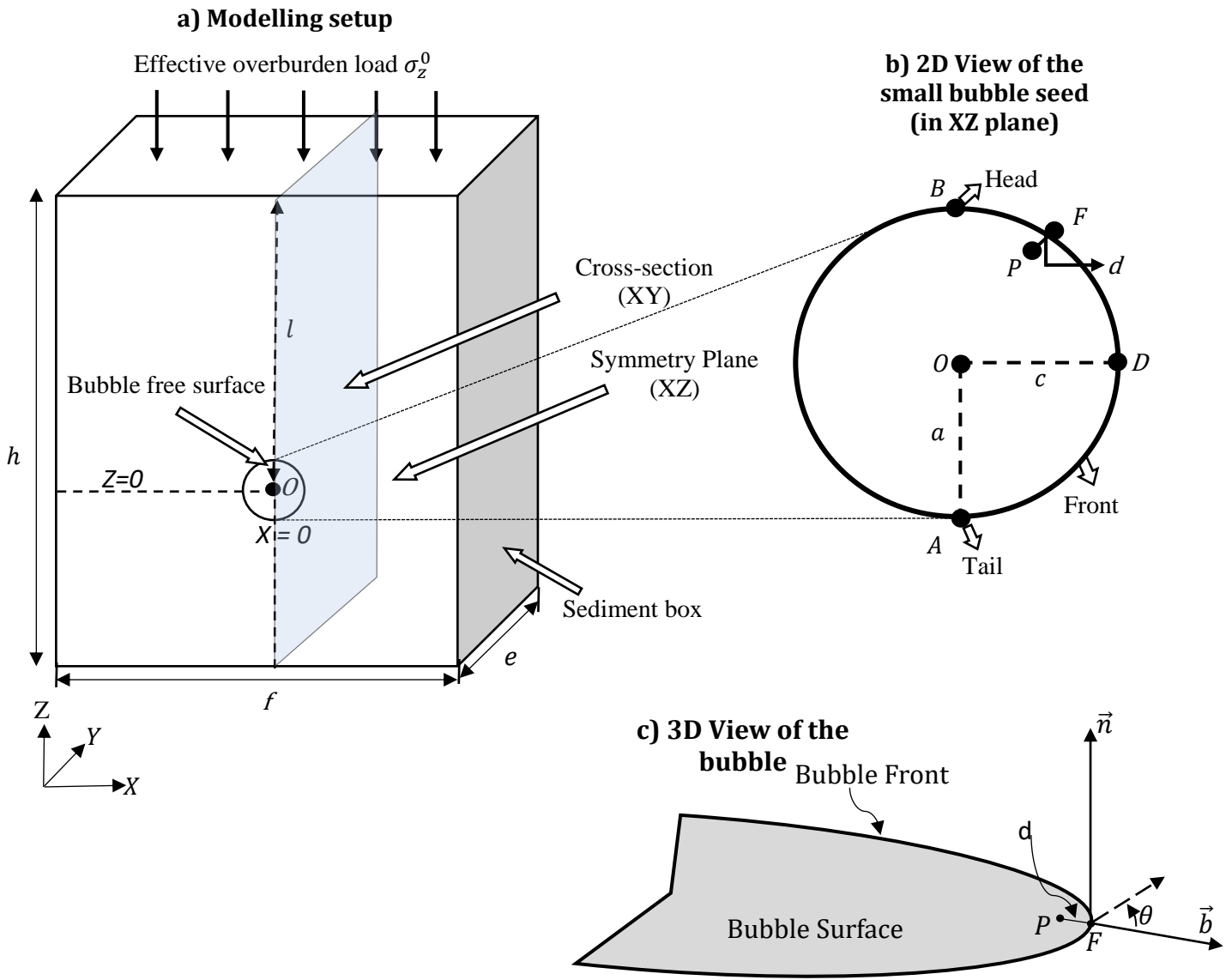

**Figure 1. a)** Geometric setup for modelling bubble growth (modified from [20]). **a)** The sediment is represented as a box. Due to symmetry, only half of this box is included in the modelling, cut by a symmetry plane, where the bubble seed is initially positioned. The dimensions of this box are *e*, *f*, and *h* (for instance, *e=0.06 m*, *f=0.23 m*, and *h=0.35 m*). The variable *l* represents the distance from the top surface of the box to the center of the bubble. An overburden load, $\sigma_z^0$, is applied on the top surface of the box. **b)** A 2D view of the initial penny-shaped bubble seed is shown. Points

*A* and *B* mark the tail and head of the bubble's front, respectively. The semi-height and semi-width of the bubble seed are designated as *a* and *c*. **c)** A 3D view of the modeled crack (bubble). The normal displacement at the crack surface (in the $\vec{n}$ direction) at point *P*, is utilized to calculate the Mode I SIF, $K_I$, at the adjacent point *F* at the bubble's front. Point *F* is located at a distance *d* along a normal direction, $\vec{b}$, to the crack front, in the local crack coordinate system.

### *2.1 Modeling the region outside the bubble*

Within the modelling framework, the transport of dissolved CH₄(aq) and the solid mechanics are comprehensively simulated throughout the bulk of the sediment cell outside the gas bubble itself (Fig.1).

*Solute transport*

The Solute Conservation Equation is modeled with the sediment, specifically for dissolved methane, $CH_4(aq)$, to accurately capture the dynamics of methane supply to a growing or migrating bubble:

$$\frac{\partial C_{CH_4}(aq)}{\partial t} + \nabla \cdot \left[-D\nabla C_{CH_4}(aq)\right] = S_{CH_4}(aq) \tag{1}$$

where $C_{CH_4}(aq)$ represents the aqueous methane concentration, $D$ is the tortuosity-corrected diffusion coefficient for CH₄, $S_{CH_4}(aq)$ denotes the rate of dissolved CH₄(aq) production or consumption (acting as a source strength), and *t* signifies time.

<u>*Boundary conditions*</u>: Given the relatively small size of the computational domain (as depicted in Figure 1), which is situated well below Sullphate-Methane Transition Zone, the variations in dissolved $CH_4(aq)$ concentrations across its boundaries are considered negligible [18]. Consequently, a uniform solute concentration is assumed across all boundaries of the domain, $C_{CH_4}(aq) = C_0$, consistent with approaches found in [23,24]. The exception is at the symmetry plane, where a no flux boundary condition is imposed, $-\vec{n}(-D\nabla C_{CH_4}(aq))=0$ (Fig.1). Here $\vec{n}$ represents the outward normal to the surface, and $D$ is the tortuosity-corrected $CH_4(aq)$ diffusion coefficient.

*Solid mechanics*

Algar et al. [30] proposed that the mechanical behavior of muddy sediment can be effectively characterized by a Kelvin-Voigt model. This model suggests a dual response: over longer timescales, the sediment behavior is primarily elastic, aligning with the principles of LEFM. However, at shorter timescales, the model incorporates an added viscosity, which accounts for time-dependent deformation. The motivation behind selecting the Kelvin-Voigt model stems from observations of bubbles within muddy sediments, which exhibit the overall mechanical stability and a lack of significant long-term deformation. If the time it takes for stress change information to propagate to the crack head to tail $\tau_f = H/U_R$ (as a function of bubble's height, $H$, and the material Raleigh velocity, $U_R$, [30]), exceeds the retardation time of Kelvin-Voigt model ($\tau_v = \mu/E$, where $\mu$ is dynamic viscosity and $E$ is Young's modulus), $\tau_v < \tau_f$, then the viscous effects become negligible and sediment responds predominantly as an elastic material. The validity of employing the LEFM has been substantiated in [21] for the typical range of kinematic viscosities observed in marine muds (e.g., between $20$ and $2000 cm^2/s$). The LEFM component is crucial for modelling two key components of bubble behavior in the current problem: the stress imposed on sediment by the elastic bubble expansion and contraction, and discrete differential fracturing of the muddy sediments by the advancing bubble front. This allows for precise tracking of the bubble's evolving shape and size.

The solid mechanics part of the model includes two primary components: elasticity and fracturing. Linear elasticity is implemented using the Force Equilibrium Equation, and models elastic deformations that arise from various forces acting on the sediment, including pressure exerted by a growing bubble, the influence of gravity and remote loads:

$$-\nabla \cdot \sigma = \vec{F_g} \tag{2}$$

where $\sigma$ is Cauchy stress tensor, $\vec{F_g}$ is gravitational load.

*Boundary Conditions:* A vertical compressive stress, $\vec{\sigma_z^0}$, is applied to the upper horizontal face of the computational domain (Fig.1), $\sigma \cdot \vec{n} = -\vec{\sigma_z^0}$, where $\sigma$ is a stress tensor. This represents the overburden load on the sediment. Zero normal displacement condition is prescribed onto all vertical faces of the domain, including the symmetry plane, $\vec{n} \cdot \vec{w} = 0$. The bottom face of the domain is fixed, $\vec{w} = 0$.

### *2.2 Modeling the bubble*

In this model, the bubble nucleation (the appearance of the crack) is not modelled, consistent with the assumptions of LEFM [14].

The conservation of methane within a growing bubble is determined by integrating the flux of the dissolved CH$_4$(aq) over the entire bubble's surface, $\alpha$ (e.g., [18]):

$$\frac{\partial(C_{CH_4}(g)V_b)}{\partial t} = \int_\alpha \vec{n} \cdot (-D\phi \nabla C_{CH_4}(aq))d\alpha \tag{3}$$

where $C_{CH_4}(g)$ and $C_{CH_4}(aq)$ represent gaseous and aqueous methane concentrations, respectively, $V_b$ is volume of the gas bubble, $\vec{n}$ signifies the outward normal at the bubble's surface $\alpha$ (Fig.1), $\phi$ represents the sediment's effective porosity, $D$ is the tortuosity-corrected methane diffusion coefficient.

The concentration of dissolved methane, $C_{CH_4}(aq)$, at the bubble's surface $\alpha$ is in equilibrium with the gaseous methane concentration inside the bubble, $C_{CH_4}(g)$. This equilibrium is governed by Henry's Law:

$$C_{CH_4}(g) = C_{CH_4}(aq)(\alpha,t) \cdot k_H \tag{4}$$

here $k_H$ represents the dimensionless Henry's Law constant [19]. The volume of the bubble at any given stage of the loading, is determined by integrating the elastic displacements, $\vec{w}$, in the direction normal, $\vec{n}$ to bubble's surface:

$$V_b = \int_\alpha \vec{n} \cdot \vec{w} d\alpha \tag{5}$$

Assuming methane behaves as an ideal gas, its concentration, $C_{CH_4}(g)$, as determined from Eq. (4), is then utilized to calculate the inner pressure of the bubble:

$$P_b = C_{CH_4}(g) R_g T \tag{6}$$

where $R_g$ denotes the gas constant, $T$ is the temperature, $\vec{n}$ represents the outward normal vector at the bubble's surface $\alpha$ (Fig.1). A continuous supply of the dissolved methane to the bubble, as described by Eq. (1) leads to a mounting inner gas bubble pressure (Eq.6). This pressure causes the bubble cavity to expand elastically, effectively preventing its closure. This inner bubble pressure, $P_b$, balances the spatial compressive stresses within the sediment that are normal to the bubble surface, which are produced by both the overburden load and gravitational force.

The inner bubble pressure, $P_b$, calculated using Eq.(6), is applied at the interface between the bubble and the surrounding sediment (i.e. at the bubble free surface, Fig.1):

$$\sigma \cdot \vec{n} = -\vec{P_b} \tag{7}$$

*Fracture modelling*

The cracking driven by increasing pressure within the bubble (e.g., Eq.6) is modeled using fundamentals of LEFM. As the bubble grows elastically (governed by Eqs.2,6,7), it gains stress intensity factor (SIF), *K*, a measure of the stress state at its front [9,14]. LEFM theory postulates that the entire behavior of a crack is solely dictated by the value of its SIF. Unlike the 2D cracks, for a simple elliptical 3D crack, the SIF is not uniform; instead it varies along the entire crack front, even when subjected to a straightforward 1D symmetric remote extension [31]. This variation in SIF along the 3D crack front is crucial for accurately predicting the bubble's complex growth pattern.

Irwin [32] established a fundamental relationship between the strain energy release rate and SIF, $G \propto K^2$, thereby linking the stress field at the crack tip to the energy balance criterion necessary for crack propagation. In the current model, the SIF at the crack (bubble) front is determined using the Displacement/Stress Extrapolation Method. This method uses the known displacement field in the immediate vicinity of the crack front, applying the general analytical solution for the crack problems [9,31]. The crack opening displacements are evaluated directly on the crack's surface (as depicted in Figure 1) using a one-point methodology. In this approach, a point *P* is defined on the crack's surface, located near a point *F* on the crack's front, at a distance *d* from *F* (Fig.1). The Mode I SIF at each point *F* along the front is then calculated using:

$$K_I = \frac{E}{4(1-v^2)} \sqrt{\frac{\pi}{2d}} 2w_n^P \tag{8}$$

where $w_n^P$ represents the projection of the displacement $\overrightarrow{w^P}$ at point *P* on the crack surface (as shown in Figure 1), to the direction $\vec{n}$ tangential to the crack front and normal to the crack surface, $\alpha$, in the local crack coordinate system; *E* denotes Young's modulus, and v is Poisson's ratio. It can be readily demonstrated that under the specific loading created by a growing bubble with inner pressure $P_b$ (as defined by Eqs.6,7), the contributions from other SIF modes are negligible, $K_{II} = K_{III} = 0$.

The direction in which a crack propagates is determined using the classical Strain Energy Density Criterion [33]. In a general loading scenario, the strain energy density factor is expressed by the following equation: $Se(\theta) = a_{11}K_I^2 + 2a_{12}K_IK_{II} + a_{22}K_{II}^2 + a_{33}K_{III}^2$, where $K_I, K_{II}, K_{III}$ are the SIFs for Mode I, Mode II, Mode III crack propagation, respectively. The coefficients $a_{ij}$ are defined by [33] as a function of the angle $\theta$. This angle $\theta$ is an out-of-plane angle measured from the direction normal to the crack's front, $\vec{b}$, within the local crack coordinate system (Fig.1c, [33,34]). Specifically, the crack surfaces α, coincide with planes where $\theta = \pm\pi$.

The theory is based on three key hypotheses:

(1) The direction of crack growth at any given point along the crack's front (for instance, point *F* in Figure 1b,c) is always oriented towards the region where the strain energy density factor is at its minimum, $Se(\theta) = Se_{min}(\theta)$, when compared to all other possible directions on the spherical surface surrounding that point;

(2) Crack extension occurs when the strain energy density factor $Se(\theta)$ in the region reaches a critical value, $Se_c(\theta)$;

(3) The variable length, $\Delta a$, of the initial crack extension is assumed to be proportional to $Se_{min}(\theta)$, such that $Se_{min}(\theta)/\Delta a = (dW/dV)_{cr}$ remains constant along the new crack front. Here, $(dW/dV)_{cr}$ represents a critical strain energy per unit volume, which is a material property (refer to Figure 6 in [31]). In our specific case with only Mode I crack propagation (i.e., $K_{II} = K_{III} = 0$), the strain energy density factor is simplified to $Se(\theta) = a_{11}K_I^2$, where the coefficient $a_{11}$ is given by $a_{11} = (1 + \cos(\theta))/16G \cdot [2 \cdot (1 - 2\nu) + 1 - \cos(\theta)]$ [33]. Here, *G* is the shear modulus, ν is Poisson's ratio. By minimizing this equation with respect to $\theta$, it can be determined that *min*($a_{11}$) is attained at $\theta = 0$. This condition ensures that the crack propagates purely in Mode I (i.e., tensile opening).

Based on point (3) above, the distribution of crack increments along the crack's front is calculated as follows [34]:

$$\Delta a = \Delta a_{max}(K_I/K_{Imax})^2 \qquad (9)$$

The fracturing occurs when the maximum SIF at the crack's front, $K_{Imax}$, exceeds a prescribed critical SIF (or sediment fracture toughness), $K_{Ic}$ (a measure of the material's resistance to tensile failure), $K_{Imax} \geq K_{Ic}$. The bubble will fracture the surrounding sediment along its front. This fracturing allows the bubble to grow and migrate upwards, which is accompanied by a temporarily drop in $K_{Imax}$ below $K_{Ic}$. However, this drop is then followed by an increase in $K_{Imax}$ due to continuous adsorption of dissolved methane $CH_4$(aq), leading to the bubble's elastic expansion (as discussed in [21,28]). The maximum crack increment (Eq.9), is consistently attained at the buoyant bubble's head (point B in Figure 1b) and progressively diminishes towards its tail (point A in Figure 1b) [21] in an isotropic and homogeneous medium (e.g., [26]). The maximum crack increment $\Delta a_{max}$ (used, for instance, in [18]) is set in our model to be significantly smaller than the size of a plastic zone at the crack tip. Following standard fracture mechanics simulation methodology, $\Delta a_{max}$ is prescribed in the model as the size of the mesh element adjacent to the crack front [34].

Once the criterion for the crack growth is met ($K_{Imax} \geq K_{Ic}$), the crack propagates in a specified direction by a calculated local increment, $\Delta a$, at its front (as defined by Equation 9). Numerically, this process is simulated through mesh deformations, where the computational mesh nodes are moved by these variable local increments $\Delta a$ at the bubble front (see Figure 1). Instead of generating an entirely new mesh at every time step, the model perturbs the existing mesh nodes within the computational domain. This ensures that the nodes conform the newly deformed boundary, which represents the bubble's free surface. To maintain mesh quality, the mesh within the domain is smoothed using the Laplace smoothing method. A complete remeshing of the entire geometry is only performed when the quality of the mesh significantly degrades.

*2.3 Numerical implementation*

The model is designed, solved and post-processed in the finite elements method (FEM)-based COMSOL Multiphysics environment. The Force Equilibrium Equation (Eq. 2) is discretized using quadratic Lagrange elements, while the Solute Conservation Equation (Eq. 1) is discretized with linear elements. These equations are solved concurrently at each time step by a PARDISO, a fully coupled direct solver. An advanced meshing strategy is developed (e.g., [20,21,26,28]) to optimize and balance the geometry resolution and computational efficiency. The finer meshes are specified at the bubble surface, its front, and their immediate surrounding within the simulation domain (as depicted in Figure 2). The simulation employs a detailed meshing strategy that begins with the finest 1D mesh elements sized at a maximum of $4 \cdot 10^{-4}$ m positioned along the initial curved bubble's front (as shown in Figure 2). These 1D elements act as the source for 2D triangular mesh elements that evolve at the bubble surfaces and adjacent symmetry plane, gradually coarsening with a growth rate of 1.2 from the edge elements. These 2D mesh elements then serve as the basis for the development of randomly oriented 3D tetrahedral mesh elements, which extend throughout the entire simulation domain (Fig.2). Initially, the model's geometry is spatially discretized using a total of 6652 tetrahedral mesh elements. Throughout the simulations, remeshing operations are performed as needed, to address the deteriorating mesh quality. The remeshings consistently adhere to the same fundamental meshing strategy outlined above. This approach ensures excellent agreement between the numerical and analytical solutions for various bubble growth parameters (e.g., [20,26,28]), while also eliminating the need for additional mesh refinement during the simulations. This, combined with implemented symmetry conditions (Fig.1) and the solver's configuration, significantly contributes to the computational efficiency of the model.

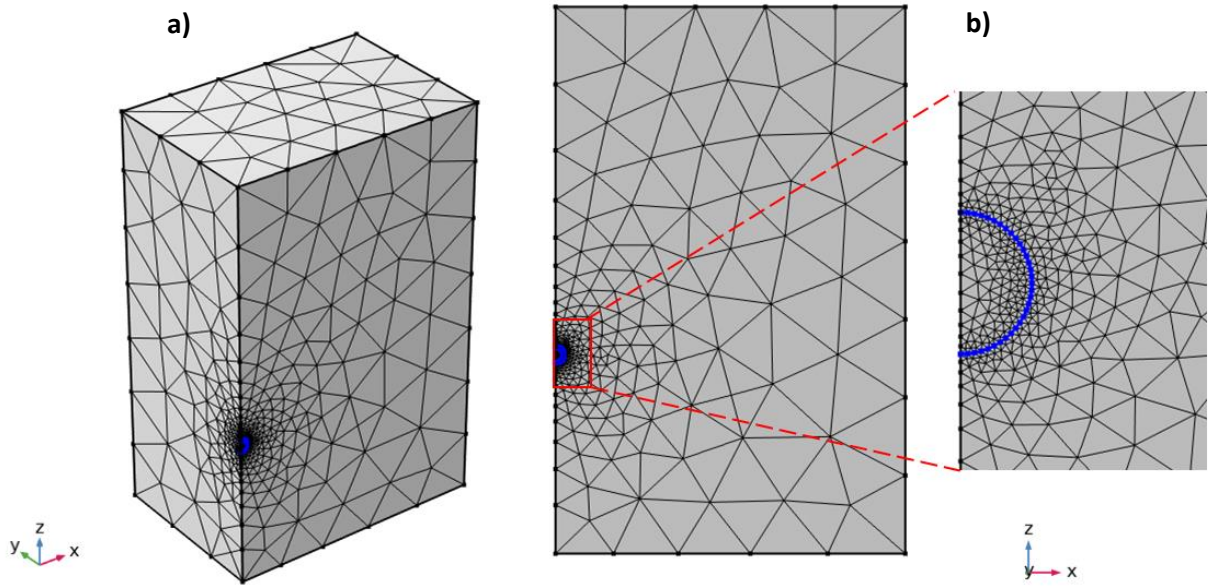

**Figure 2. a)** Randomly oriented 3D tetrahedral mesh elements developed over the entire computational domain. **b)** 2D frontal view on the triangular meshes developed at and around the bubbl's surface (including the inset with magnification) that serve as an origin for the 3D mesh evolution (in **a**), see the text for more meshing details.

*2.4 Simulation conditions*

A computational geometry is prescribed with the following parameters that may vary depending on the specific simulation being performed (Fig.1): *h=0.35 m, f=0.23 m, e=0.06* m. The computation itself initiates with a small flat penny-shaped bubble having an initial radius of *R=4 mm*. This particular size is chosen to ensure a minimal difference in the SIF between the tail (point A) and the head (point B) of the crack/bubble front (Fig.1, [21]), once the bubble inflates. Any bubble growth occurring prior to this 4 mm radius stage, is not considered in the simulation [1].

Input data for the simulations (Table 1) are based on the summer conditions at the NRL site in Eckernförde Bay (26 m water depth) at 1 m sediment depth.

**Table 1**. Summary of geochemical and mechanical data used as input for the simulations (for justifications of these parameters, please refer to [28])

| Parameter | Value |
|---|---|
| Dissolved aqueous pore space methane concentration ($C_{CH_4}(\text{aq})$) | 0.1 kg m$^{-3}$ |
| Molecular diffusion coefficient of methane in free-solution ($D_m$) | 10$^{-9}$ m$^2$ s$^{-1}$ |
| Sediment bulk density ($\rho$) | 1240 kg m$^{-3}$ |
| Effective overburden load ($\sigma_z$) | 5 kPa |
| Sediment effective porosity ($\phi$) | 0.2 |
| Sediment Young's modulus ($E$) | 5.5·10$^5$ Pa |
| Sediment Poisson's ratio (v) | 0.45 |

3. **Method validation**

The model was extensively verified against the available analytical solutions for the growing bubble parameters (e.g., [20,26,28]). The bubble shape and size are fundamentally determined by the elastic and fracture properties of muds, defined by the LEFM theory: i.e., Young's modulus, *E*, Poisson's ratio, $v$, and Mode I fracture toughness, *K$_{Ic}$*. Building on this theoretical framework, analytical solutions have been developed for key characteristics of mature elliptic bubbles, including their semi-height, $a$, and normal opening (COD –crack opening displacement) (e.g., [20,28,35]):

$$a = \left(-\frac{K_{Ic}[(1-2k^2)E(k)-k'^2K(k)]}{2\sqrt{\pi}k^2k'\frac{v}{1-v}\rho g}\right)^{2/3} \quad (10)$$

where $v$ is Poisson's ratio, $K_{Ic}$ is Mode I fracture toughness, $\rho$ is the bulk sediment density, $g$ is gravitational acceleration, $a$ and $c$ are major and minor semi-axes of the elliptical bubble (Fig.1b), $k' = c/a$, $k^2 = 1 - (c/a)^2$, and $K(k)$ and $E(k)$ are complete elliptic integrals of the first and second kind, respectively.

The COD at any point of the surface of a mature bubble (with a closed tail) is derived as:

$$COD = \frac{2(1-v^2)\sqrt{a}\sqrt{1-r^2}}{E}\left[\frac{K_{Ic}+K_{Ic}\cos(\varphi)}{\sqrt{\pi}}\right] \quad (11)$$

where $r$ is a non-dimensional radius $r = \sqrt{z^2/a^2 + x^2/c^2}$, $\varphi$ is a parametric elliptical angle defined for the points of the crack front: $\varphi = 0$ is at the bubble head and $\varphi = \pi$ is at the bubble tail, $z = a \cdot \cos(\varphi)$, and $x = c \cdot \sin(\varphi)$ (see [20] for more details).

The shape and size of the mature bubbles modeled numerically and analytically (Eqs.10,11), were compared (Fig. 3). This validation conducted across varying fracture toughness, consistently shows a good agreement between the modelled and analytically derived results.

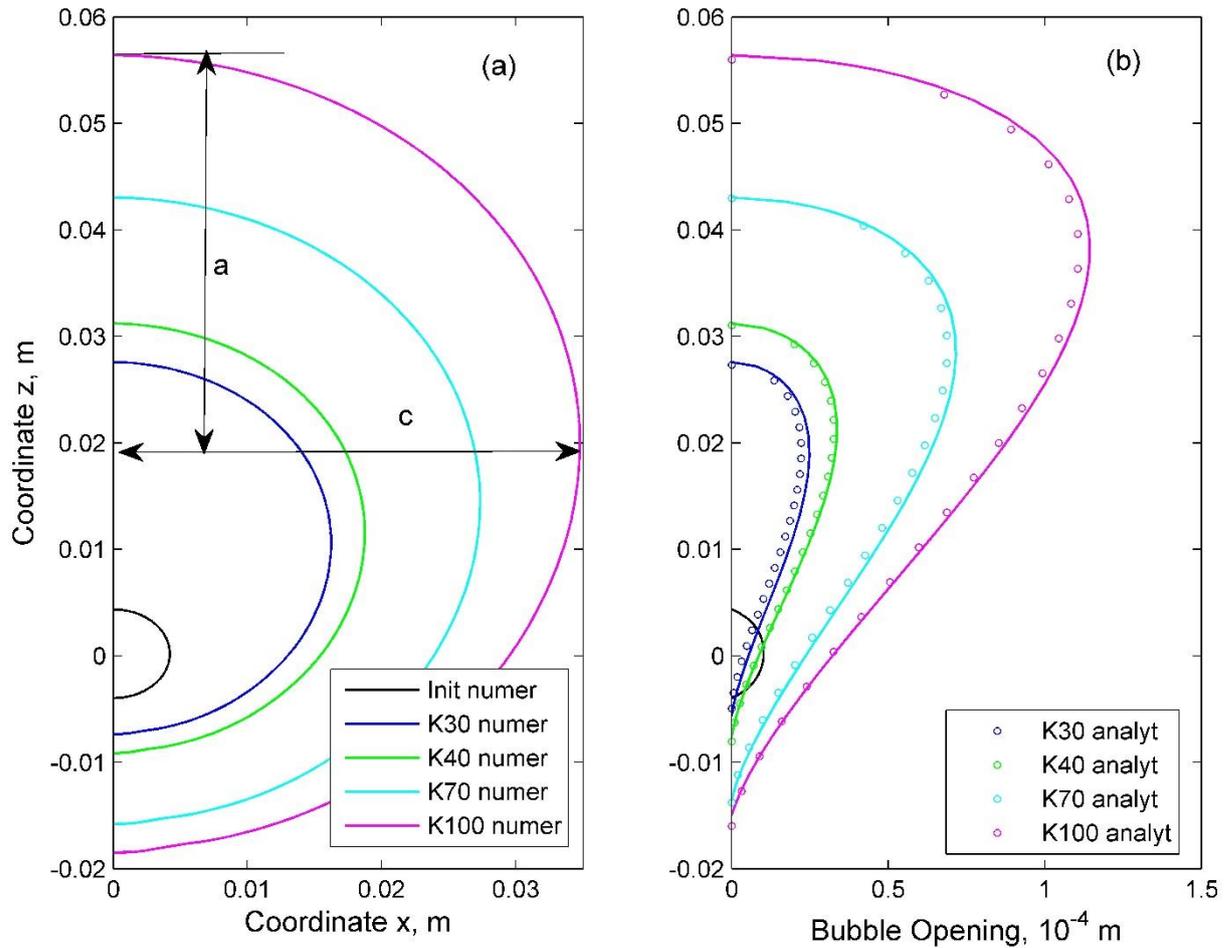

**Figure 3.** Shape and size of the initial penny-shaped and mature bubbles (with a closed tail) modeled numerically and analytically across varying fracture toughness, $K_{Ic}$ (adopted from [20]). For clarity, only one half of the symmetric structure is shown (refer to Figures 1,2 and source [20]). **a)** Bubbles' front view is depicted. All the modeled bubbles exhibit an elliptical shape. They maintain an approximate constant planar inverse aspect ratio $k' = c/a \cong 0.92$, where $a$ and $c$ represent the major and manor semi-axes of the ellipse, respectively (Fig.1). **b)** Bubbles' cross-section along the Z axis at X=0 (Fig.1). Initial penny-shaped bubble displays a symmetric elliptic opening (indicated by the black line). All the mature bubbles formed under the different fracture toughness, $K_{Ic}$, conditions, consistently develop into an "inverted tear drop" shape. Sediments with lower fracture toughness tend to produce smaller flatter bubbles. Conversely, higher fracture toughness results in larger and more significantly inflated bubbles. The results obtained from both the analytical

solutions (Eqs.10,11, represented by circles) and the numerical modeling (represented by solid lines) show a good agreement across all tested values of fracture toughness.

4. Limitations

The present model is applicable to individual bubbles. Currently, there isn't comprehensive theory, for upscaling the entire set of the bubble characteristics (like bubble shapes, sizes, and orientations, as seen in Fig.3), to the effective mechanical and physical properties of gassy aquatic muds, which are crucial for remote sensing. Therefore, this model will serve as a foundational tool for developing accurate upscaling methods. These methods will further preserve the essential details of single bubble behavior, including growth physics, and controlling factors. This will ultimately improve the accuracy of the acoustic applications. The proper upscaling is vital for assessing and predicting sediment destabilization and geohazard, and it could also help resolve the long-standing uncertainty related to the net $CH_4$ fluxes from aquatic muds.

**Author statement**

**Regina Katsman**: Conceptualization, Methodology, Software, Validity tests, Writing- Original draft preparation, Visualization, Investigation.

**Acknowledgments**

This work was supported by the Israel Science Foundation, grant No. 1441-14, and by the U.S.-Israel Binational Science Foundation, grant No. 2018150.

**Declaration of interests**

☒ The authors declare that they have no known competing financial interests or personal relationships that could have appeared to influence the work reported in this paper.

☐ The authors declare the following financial interests/personal relationships which may be considered as potential competing interests: